\documentclass[intlimits,twoside,a4paper]{article}

\usepackage{amsmath,amssymb}
\usepackage{graphicx}

\usepackage[T2A]{fontenc}
\usepackage[cp1251]{inputenc}
\usepackage{multirow}

\usepackage[eqsecnum]{cmpj2}

\issue{2016}{19}{1}{13801}
\doinumber{10.5488/CMP.19.13801}

\title[Modelling and measurements of fibrinogen adsorption on positively charged microspheres]%
{Modelling and measurements of fibrinogen adsorption on positively charged microspheres}
\author[P. \.Zeliszewska \textsl{et al}.]{P. \.Zeliszewska\refaddr{label1},
A. Bratek-Skicki\refaddr{label1}, Z. Adamczyk\refaddr{label1}, M. Cie\'sla\refaddr{label2}}
\addresses{
\addr{label1} J. Haber Institute of Catalysis and Surface Chemistry Polish Academy of Sciences, \\ Niezapominajek 8, 30-239 Cracow, Poland
\addr{label2} M. Smoluchowski Institute of Physics, Jagiellonian University, \L{}ojasiewicza 11, 30-348 Cracow, Poland
}

\authorcopyright{P. \.Zeliszewska, A. Bratek-Skicki, Z. Adamczyk, M. Cie\'sla, 2016}

\date{Received October 28, 2015, in final form December 3, 2015}

\begin{document}

\maketitle

\begin{abstract}
Adsorption of fibrinogen on positively charged microspheres was theoretically and experimentally studied. The structure of monolayers and the maximum coverage were determined by applying the experimental measurements at $\text{pH} = 3.5$ and 9.7 for NaCl concentration in the range of 10$^{-3}- 0.15$~M. The maximum coverage of fibrinogen on latex particles was precisely determined by the AFM method. Unexpectedly, at $\text{pH} = 3.5$, where both fibrinogen molecule and the latex particles were positively charged, the maximum coverage varied between 0.9~mg~m$^{-2}$ and 1.1~mg~m$^{-2}$ for 10$^{-2}$ and 0.15~M NaCl, respectively. On the other hand, at $\text{pH} = 9.7$, the maximum coverage of fibrinogen was larger, varying between 1.8~mg~m$^{-2}$ and 3.4~mg~m$^{-2}$ for $10^{-2}$ and 0.15~M NaCl, respectively. The experimental results were quantitatively interpreted by the numerical simulations.
\pacs 82.70.Dd, 87.15A, 87.14.E
\end{abstract}

\section{Introduction}

Protein adsorption on various surfaces has received a considerable attention due to its importance in biomedical fields. New biotechnological methods of protein production, purification and separation depend on their interfacial properties. Furthermore, the protein adsorption at solid/liquid interfaces enabled the development of diverse biomedical applications, such as biosensors, immunological tests, and drug delivery systems. On the other hand, in biomaterial field, protein adsorption is undesirable because it can cause adverse responses such as blood coagulation and complement activation \cite{b1}.

Fibrinogen (Fb) is a major serum blood protein that plays an essential role in the clotting cascade initiated by thrombin. Fibrinogen molecules exhibit a strong tendency to adsorb on various surfaces under broad range of conditions \cite{b2,b3} mediating cellular interactions that are a key event in determining biocompatibility of these materials \cite{b4}. On the other hand, fibrinogen monolayers adsorbed on various synthetic materials may promote platelet adhesion that often leads to fouling of artificial organs~\cite{b6,b7}.

Due to its fundamental significance, the chemical structure of fibrinogen and its bulk physicochemical properties have been extensively studied \cite{b8,b9,b10,b11,b12,b13,b14,b15}. It was established \cite{b9} that the fibrinogen molecule is a symmetric dimer composed of three identical pairs of polypeptide chains, referred to as A$\alpha$, B$\beta$ and $\gamma$ chains \cite{b10}. They are coupled in the middle of the molecule through a few disulfide bridges forming a central E nodule. The longest A$\alpha$ chain is composed of 610 amino acids, the B$\beta$ chain comprises 460 amino-acids and the $\gamma$ chain 411 amino acids. Accordingly, the molar mass of the fibrinogen molecule is equal to 338~kDa \cite{b11}. It is interesting to mention that a considerable part of the A$\alpha$ chains extends from the core of the molecule forming two polar appendages (arms) called the $\alpha$C domains \cite{b12}.

Information on fibrinogen's molecule shape and dimensions was derived from electron \cite{b13,b14,b15,b16} and atomic force microscopy (AFM) \cite{b17,b18,b19,b20,b21,b22}. It was established that the molecule has a co-linear, trinodular shape with a total length of 47.5~nm \cite{b13}. The two equal end domains of rather irregular shape are approximated by spheres having a diameter of 6.5~nm; and the middle domain has a diameter of 5~nm. Additionally, in references \cite{b23,b24} it was predicted that the length of the side arms is equal to 18~nm and that they are positively charged in the $\alpha$C domains for the broad range of pHs from 3.5 to 9.7.

Due to its significance, the adsorption of fibrinogen at solid/electrolyte interfaces has been investigated by various experimental methods such as ellipsometry \cite{b25}, Total Internal Reflection (TIRF) \cite{b26}, AFM \cite{b27}, radiolabeling \cite{b28}, Quartz Crystal Microbalance (QCM) \cite{b29}, streaming potential \cite{b23}, etc. However, only in the  work of Brash et al. \cite{b4} positively charged surfaces were investigated, prepared by controlled adsorption of polycations on a glass substrate. The radiolabeled fibrinogen was used to monitor the kinetics of adsorption. It is postulated that three populations of fibrinogen appear in the adsorbed monolayers: non- exchanging, slowly, and rapidly exchanging. It was shown that for positively charged surfaces, the amount of fast and slow exchanging molecules is oppositely different from the neutral and negatively charged surfaces.

In contrast to the vast literature dealing with other proteins, only a few studies focused on fibrinogen adsorption on polymeric microspheres have been reported \cite{b30,b31}. In  reference \cite{b32}, the effect of ionic strength in the adsorption of fibrinogen on polystyrene  microspheres (latex) at $\text{pH} = 7.4$ was studied. The electrophoretic mobility measurements were performed to control the progress of protein adsorption under in situ conditions. The coverage of the protein was determined by a concentration depletion method involving AFM imaging of residual fibrinogen. It was revealed that the maximum coverage of fibrinogen on latex varied between 1.9 and 3.2~mg~m$^{-2}$ for $10^{-3}$ and 0.15~M NaCl concentration, respectively. In a recent publication \cite{b31}, adsorption of fibrinogen at positively charged amidine microspheres was studied at pH of 7.4. Quite unexpectedly, the maximum coverage of the protein was much smaller than for negatively charged latex, equal to 0.6 and 1.3~mg~m$^{-2}$ for $10^{-3}$ and 0.15~M NaCl, respectively. This anomalous result was interpreted in terms of the random sequential adsorption model, by postulating a side-on adsorption of fibrinogen molecules at the latex surface.

The main goal of this work is to systematically study fibrinogen adsorption at positively charged microspheres at $\text{pH} = 3.5$ and 9.7 in order to quantitatively determine the maximum coverage as a function of this parameter. These results, supplemented by electrophoretic mobility measurements furnishing the charge density data, allow one to elucidate fibrinogen adsorption mechanisms for a broad range of pH that has a significance for basic science. This is also of a practical interest for developing a robust procedure of preparing stable fibrinogen monolayers at latex particles of a well-controlled coverage and a known orientation of molecules. Such latex/fibrinogen complexes can be exploited to study interactions with other proteins (antibodies) and low molar mass ligands.

\section{Materials and methods}

\subsection{Experimental}

Fibrinogen from human blood plasma, fraction I, type IV used for our study was purchased from Sigma (F4753) in the form of a powder. The sample contains 65\% protein, 15\% sodium citrate, and 25\% sodium chloride. The dynamic surface tension method was used to check the purity of fibrinogen solution. The bulk concentration of the fibrinogen in electrolyte solutions was determined by the BCA method (Bicinchoninic acid Protein Assays) \cite{b33}.

High-Performance Liquid Chromatography (HPLC) experiments were performed using an Knauer system. The column was packed with a composite of cross-linked agarose and dextran. The flow rate was 0.5~ml/min. The absorptiometric detection was monitored at 280~nm.

The water which was used in all experiments was purified by the Milipore Elix 5 instruments. All other reagents were purchased from Sigma--Aldrich and used without any purification.

Amidine microspheres (latex)  used in our measurements were purchased from Invitrogen. This latex was positively charged, surfactant free with concentration equal to 3.7\% and nominal size of 800~nm.
The pH of Fb solutions and latex suspension was regulated within the range by the addition of HCl or NaOH. Buffers were not used for experiments, due to their specific adsorption on monolayers.

The dynamic light scattering (DLS) was used to determine diffusion coefficients of fibrinogen and latex particles. On the other hand, the electrophoretic mobility of fibrinogen  and fibrinogen-covered microspheres was measured by the Laser Doppler Velocimetry (LDV) technique. The diffusion coefficient and the elctrophoretic mobility were measured using the Zetasizer Nano ZS Malvern instruments.
The concentration depletion method described in reference \cite{b34} was used to determine the excess of fibrinogen after adsorption on the latex. This method consisted of several stages: the experiment was started by transferring the fibrinogen latex mixture after the adsorption step to the diffusion cell. Then, a few mica sheets were immersed in the suspension for 30 minutes (adsorption time). Subsequently, the fibrinogen covered mica sheets were rinsed using a pure electrolyte with the same pH and ionic strength as for the adsorption of fibrinogen on latex particles. The AFM imaging was used to determine the average number of fibrinogen molecules adsorbed over equal sized surface areas randomly selected over the mica sheets.

\subsection{Theoretical modelling}

The modelling of fibrinogen monolayer formation on microparticle surfaces  was carried out by applying the random sequential adsorption (RSA) approach developed in references \cite{b35,b36} for quantifying irreversible adsorption proteins (ferritin) on flat interfaces. In these calculations, the specific interactions among protein molecules were neglected and their shape was approximated by a circular disk. Later on, the RSA model was extensively used for calculating the kinetics, the maximum (jamming) coverage and the monolayer structure of non-spherical particles of various shapes \cite{b37,b38,b39,b40}. However, all these results were obtained for  hard (non-interacting) particles and flat interfaces of infinite extension,  by neglecting the curvature effect that can influence both the structure and the maximum coverage of monolayers.

The general rules of the Monte-Carlo simulation scheme based on the RSA approach are as \linebreak follows~\cite{b36,b38}:

(i) a virtual particle (molecule) is created, whose position within the simulation domain and orientation are selected at random with a probability depending on the interaction energy,

(ii) if the particle fulfills the pre-defined adsorption criteria it becomes irreversibly deposited and its position remains unchanged during the entire simulation process,

(iii) if the deposition criteria are violated, a new adsorption attempt is made, uncorrelated  with the previous attempts.

\begin{figure}[!b]
\centerline{
\includegraphics[width=0.6\textwidth]{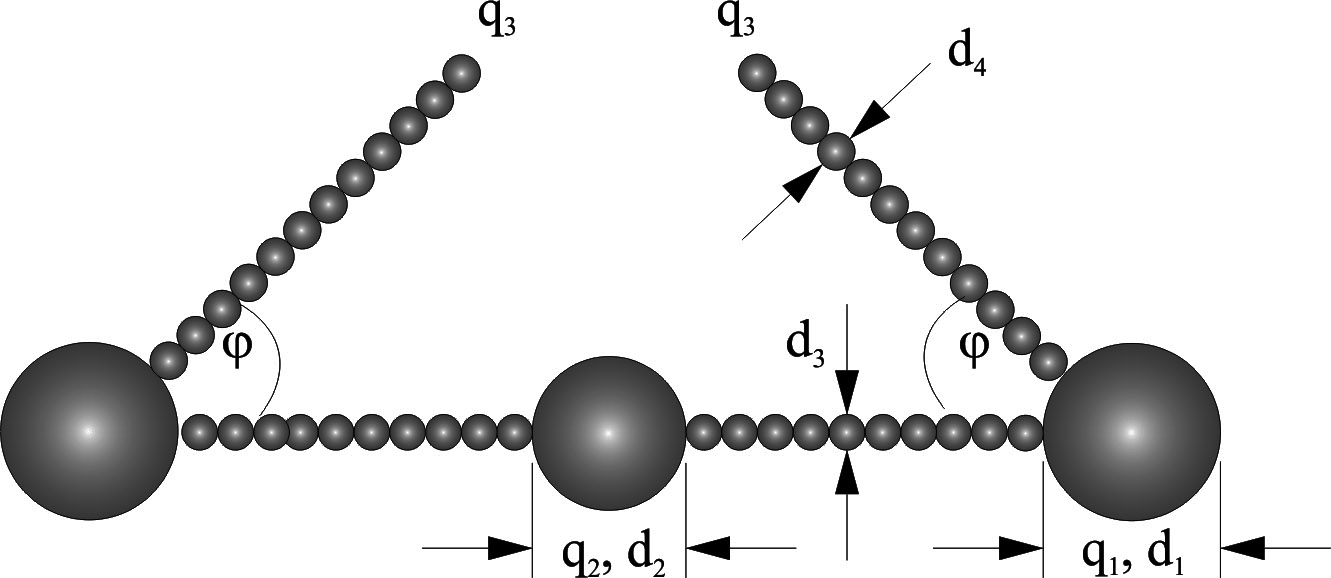}
}
\caption{The  bead model of the fibrinogen molecule \cite{b24}, $d_1 = 6.7$~nm, $d_2 = 5.3$~nm, $d_3 = d_4 = 1.5$~nm.
$\varphi = 56^{\circ}$ ($\text{pH} = 7.4$).}
\label{fig1}
\end{figure}

Usually, two major deposition criteria are defined: (i) there should be no overlapping of any part of the virtual particle with the previously adsorbed particles and (ii) there should be a physical contact of the particle with the interface.
 Despite the simplicity of the governing rules, the RSA modelling is a powerful tool for efficiently producing populations of a large  number of molecules $N_p$ (often exceeding
$10^5$). It is also flexible because  adsorption processes of   anisotropic particles at interface of various geometry, for example on spherical microparticles, can be efficiently treated. In this work, adsorption of fibrinogen on microparticle surfaces is simulated by using the bead Model B (see figure~\ref{fig1}), previously used for describing adsorption at flat and negatively charged interfaces \cite{b23}. In this model, the shape of the fibrinogen molecule is approximated by a string of 23 co-linear touching spheres of various diameters. The two external spheres have diameters of 6.7~nm and the central sphere has a diameter of 5.3~nm. The side arms are approximated as straight sequences of 12 beads of equal size, having also the diameter of 1.5~nm. These side chains form the angle $\varphi$ with the core part of the fibrinogen molecule.  Electric charges denoted by $q_1$, $q_2$, $q_3$ (see table~\ref{tab0}) are attributed to various parts of the fibrinogen molecule as shown in figure~\ref{fig1}. The magnitude of these charges can be estimated from the electrophoretic mobility measurements as discussed later on.

\begin{table}[!t]
\caption{Model parameters used in numerical simulations. Diameters of the spheres: $d_1 = 6.7$~nm, $d_2 = 5.3$~nm, $d_3 = d_4 = 1.5$~nm were the same for all pH's and ionic strengths.}
\label{tab0}
\vspace{2ex}
\begin{center}
\begin{tabular}{|c|c|c|c|c|c|c|c|c|}
\hline\hline
\multirow{3}{*}{pH} & Ionic & $q_1$ & $q_2$ & $q_3$ & $\phi$ \\
& strenght [M] & [$e$] & [$e$] & [$e$] & deg \\
\hline
\hline
\multirow{2}{*}{3.5}
 & 10$^{-2}$ & --6 & 0 & 0 & \multirow{2}{*}{95} \\
 \cline{2-5}
 & 0.15      & --8 & 0 & 0 & \\
\hline
\multirow{2}{*}{9.7}
 & 10$^{-3}$ & --10 & 0 & 0 & \multirow{3}{*}{30} \\
 \cline{2-5}
 & 10$^{-2}$ & --20 & 0 & 0 & \\
 \cline{2-5}
 & 0.15      & --31 & 0 & 0 & \\
\hline\hline
\end{tabular}
\end{center}
\end{table}

The lateral electrostatic interactions between the adsorbed fibrinogen molecules were accounted for by using the Yukawa (screened Coulomb) pair potential $\phi$ given by
\begin{equation}
\phi(r_{ij}) = \frac{q_i q_j}{4\pi \epsilon r_{ij}} \exp\left[ {-\kappa \left(r_{ij} - \frac{d_i}{2} - \frac{d_j}{2}\right) }\right],
\label{eq1}
\end{equation}
where $q_i$, $q_j$ are the effective (electrokinetic) charges of the two beads $i$ and $j$, $r_{ij}$ is the distance between the bead centers having the diameters $d_i$ and $d_j$, and $\kappa^{-1} = \left( {\epsilon k_\text{B} T}/{2 e^2 I} \right)^{1/2}$ is the electrical double-layer thickness, $\epsilon$ is the permittivity of the medium, $k_\text{B}$ is the Boltzmann constant, $T$ is the absolute temperature, $e$ is the elementary charge and $I$ is the ionic strength.

In order to calculate the net interaction between the molecules, the pair potentials given by equation (\ref{eq1}) were evaluated over the interaction area, significantly exceeding the double-layer thickness (by excluding interaction of the beads belonging to the same fibrinogen molecule).

The model molecules were adsorbed according to the above RSA scheme on a spherically shaped interface whose radius of curvature exactly matched the dimension of the latex particles used in adsorption experiments. The electrostatic interactions of the protein with the latex surface were assumed to be of the square well (perfect sink) type.

Typically, in one simulation run, 1500 molecules were generated. Therefore, in order to improve the statistics, the averages from ca. 50 independent runs were taken, with the total number of particles exceeding 70\,000. This ensures a relative precision of the simulation better that 0.5\%. The primary parameter derived from these simulations was the average number of molecules adsorbed on latex particles calculated as a function of time. By extrapolating this dependence to infinite time using the procedure previously described \cite{b40} one obtains the maximum number of molecules $N_p$ adsorbed on the latex particles. Consequently, the surface concentration of the protein is $N_p / \Delta S$ (where $\Delta S = \pi d_\text{l}^2$ is the surface area of the latex particle of the diameter $d_\text{l}$) and the dimensionless coverage is calculated from the formula
\begin{equation}
\Theta = \frac{S_g N_p}{\Delta S},
\label{eq2}
\end{equation}
where $S_g$ is the characteristic cross-section area of the protein molecule.

Knowing $N_p$, the  dimensional coverage commonly used for the interpretation of experimental data is calculated from the following dependence:
\begin{equation}
\Gamma_\text{f} =  \frac{M_w}{A_v} N_p,
\end{equation}
where $M_w$ is the molar mass of fibrinogen and $A_v$ is the Avogadros's constant.

\section{Results and discussion}

\subsection{Fibrinogen and latex characteristics in the bulk}

As mentioned before, the diffusion coefficient of the microspheres was measured by the DLS for various ionic strengths and pHs. Knowing the diffusion coefficient, the hydrodynamic diameter $d_\text{l}$ [nm] was calculated using the Stokes-Einstein relationship.  In this way, the hydrodynamic diameter of microspheres was determined to be $860 \pm 15$~nm, for $10^{-3}$~M, $830 \pm 10$~nm, for $10^{-2}$~M, and $815 \pm 10$~nm for 0.15~M.

The electrophoretic mobility $\mu_\text{e}$ of  microspheres was determined by using the laser Doppler velocimetry (LDV) technique. Zeta potential values of latex particles for ionic strength $10^{-3}$~M, $10^{-2}$~M, 0.15~M NaCl and $\text{pH} = 3.5$ are equal to 78~mV, 87~mV and 40~mV, respectively. Knowing the zeta potential of latex particles, one can calculate its uncompensated charge using the Gouy-Chapman relationship for a symmetric 1:1 electrolyte \cite{b34}
\begin{equation}
\sigma_0 = \frac{\left( 8\epsilon k_\text{B} T n_b \right)^{\frac{1}{2}}}{0.160} \sinh \left( \frac{e \zeta}{2k_\text{B} T}\right),
\label{eq4}
\end{equation}
where $\sigma_0$ is the electrokinetic charge density of latex particles expressed in
$e$~nm$^{-2}$ and $n_b$ is the number concentration of the salt (NaCl) expressed in m$^{-3}$.

By using the experimental zeta potential values one obtains for $\text{pH} = 3.5$ from equation (\ref{eq4}) $\sigma_0 = 0.057$, 0.19 and 0.25 $e$~nm$^{-2}$ for the NaCl concentration of $10^{-3}$, $10^{-2}$ and 0.15~M, respectively.  Analogously,  for $\text{pH}=9.7$, $\sigma_0 = 0.042$ and 0.14 for the NaCl concentration of $10^{-3}$ and 0.15~M, respectively (see table~\ref{tab1}).

\begin{table}[!h]
\caption{Electrophoretic mobility, zeta potential (calculated from the Henry's model) and charge density of latex and human serum fibrinogen.}
\label{tab1}
\vspace{2ex}
\begin{center}
\begin{tabular}{|c|c|c|c|c|c|c|c|c|}
\hline\hline
\multirow{3}{*}{pH} & Ionic & \multicolumn{3}{|c|}{Latex A800} & \multicolumn{4}{|c|}{Human serum fibrinogen} \\
\cline{3-9}
& strength & $\mu_\text{e}$ & $\zeta_\text{l}$ & $\sigma_0$ & $\mu_\text{e}$ & $\zeta_\text{f}$ &  \multirow{2}{*}{$N_\text{c}$*} &  \multirow{2}{*}{$N_\text{c}$**} \\
& [M] & [\textmu m cm / Vs] & [mV] & [nm$^{-2}$] & [\textmu m cm / Vs] & [mV] &  &  \\
\hline
\hline
\multirow{3}{*}{3.5}
 & $1.3\times10^{-3}$ & 5.8 & 78 & 0.057 & 1.4  & 25  & 15  & 33 \\
 \cline{2-9}
 & 10$^{-2}$     & 6.7 & 87	& 0.19  & 0.94 & 16  & 10  & 42 \\
 \cline{2-9}
 & 0.15          & 3.1 & 40 & 0.25  & 0.52 & 9.0 & 5.7 & 63 \\
\hline
\multirow{3}{*}{7.4$^\dagger$}
 & 10$^{-3}$ & 5.3 & 71 & 0.044 & --0.94 & --18 & --10 & --21 \\
 \cline{2-9}
 & 10$^{-2}$ & 6.2 & 81 & 0.17  & --0.56 & --9.8 & --6.2 & --25 \\
 \cline{2-9}
 & 0.15      & 2.6 & 34 & 0.20  & --0.30 & --4.4 & --3.3 & --37 \\
\hline
\multirow{3}{*}{9.7}
 & 10$^{-3}$ & 5.2 & 70 & 0.042 & --0.91 & --17  & --10  & --20 \\
 \cline{2-9}
 & 10$^{-2}$ & 5.5 & 72 & 0.14  & --0.91 & --16  & --10  & --40 \\
 \cline{2-9}
 & 0.15      & 1.8 & 24 & 0.14  & --0.51 & --7.6 & --5.6 &	--62 \\
\hline\hline
\end{tabular}
\end{center}
\small{Footnotes: \\
*$N_\text{c} = Q_\text{c}^0 / 1.602 \times 10^{-19}$,\qquad
**$N_\text{c} = Q_\text{c} / 1.602x\times 10^{-19}$,\qquad
$^\dagger$ Previous results obtained in reference \cite{b31}.}
\end{table}

The electrophoretic mobility of fibrinogen molecules was determined previously \cite{b31}. These data are also given  in table~\ref{tab1}. As can be seen, the electrophoretic mobility of fibrinogen was positive for $\text{pH} = 3.5$ and negative for $\text{pH} = 7.4$ and 9.7.  By using the electrophoretic data, one can calculate the electrokinetic charge of the fibrinogen molecule $Q_\text{c}^0$ (expressed in Coulombs) from the Lorenz-Stokes relationship \cite{b23,b41}
\begin{equation}
Q_\text{c}^0 = 3 \pi d_\text{H} \eta \mu,
\label{eq5}
\end{equation}
where $d_\text{H}$  is the hydrodynamic diameter of fibrinogen, $\eta$ is the dynamic viscosity of the solvent (water).

It should be mentioned that equation (\ref{eq5}) is valid for molecules of arbitrary shape but its accuracy is limited for higher ionic strengths. Therefore, in this case, the following equation was used, valid for an arbitrary ionic strength and spherical particles \cite{b34}
\begin{equation}
Q_\text{c} = \frac{2}{3} Q_\text{c}^0 \frac{1+\kappa d_\text{H}}{f_\text{H} (\kappa d_\text{H})},
\label{eq6}
\end{equation}
where $f_\text{H}(\kappa d_\text{H})$ is the Henry's function.

The results calculated from equations (\ref{eq4})--(\ref{eq6}), converted to the number of charges $N_\text{c}$ are collected in table~\ref{tab1}. As can be noticed, at pH of 3.5, fibrinogen molecule acquires  a net  positive charge, whereas at higher pHs of 7.4 and 9.7, the charge is highly negative. It should also be mentioned that from the electrophoretic mobility  measurements alone  one cannot predict  in a unique way the charge distribution among various parts of fibrinogen molecule.  However, from the  diffusion  coefficient and dynamic viscosity measurements reported in references \cite{b24} it is predicted that the positive charge is concentrated in the end parts of the side arms, which means that $q_3$ remains positive for pH range 3.5--9.7. Accordingly,  $q_1$ and $q_2$ should be negative. However, a precise charge distribution at this pH is not known. This can only be empirically determined form the thorough  adsorption experiments reported below.

\subsection{Fibrinogen adsorption on microspheres}

Fibrinogen adsorption on polymeric microspheres was monitored by measuring the changes in electrophoretic mobility (zeta potential) induced by this process. The steps of the experiment were as follows:

\begin{itemize}
\item[(i)] measurement of the zeta potential of bare latex particles whose concentration $c_\text{l}$ was equal to 60~mg~L$^{-1}$,

\item[(ii)] adsorption of fibrinogen on latex particles by filling the cell with the  fibrinogen solution of an opportune concentration $c_\text{f}$ (0.1--5~mg~L$^{-1}$) for 600 seconds,

\item[(iii)] purification of the latex suspension by using a membrane filtration and measurement of the electrophoretic mobility of latex using the electrophoretic method.

\end{itemize}

This procedure is reproducible and allows one to determine the changes of zeta potential as a function of fibrinogen concentration added to the latex suspension.

The fibrinogen coverage on latex particles was calculated by using the following formula
\begin{equation}
\label{eq7}
\Gamma_\text{f} = 10^{-3} \frac{v_\text{s} c_\text{f}}{S_\text{l}},
\end{equation}
where $\Gamma_\text{f}$  is the fibrinogen coverage on microspheres  (expressed in mg~m$^{-2}$), $v_\text{s}$ is the volume of suspension mixture, $c_\text{f}$ is the initial concentration of fibrinogen in the suspension (after mixing with the latex), and $S_\text{l}$ is the surface area of latex expressed in m$^2$, given by $S_\text{l} = 6 c_\text{l} v_\text{s} / d_\text{l} \rho_\text{l}$, where $\rho_\text{l}$ is the latex specific density, equal to $1.05 \times 10^3$~kg~m$^{-3}$.

\begin{figure}[!t]
\begin{center}
\includegraphics[width=0.45\textwidth]{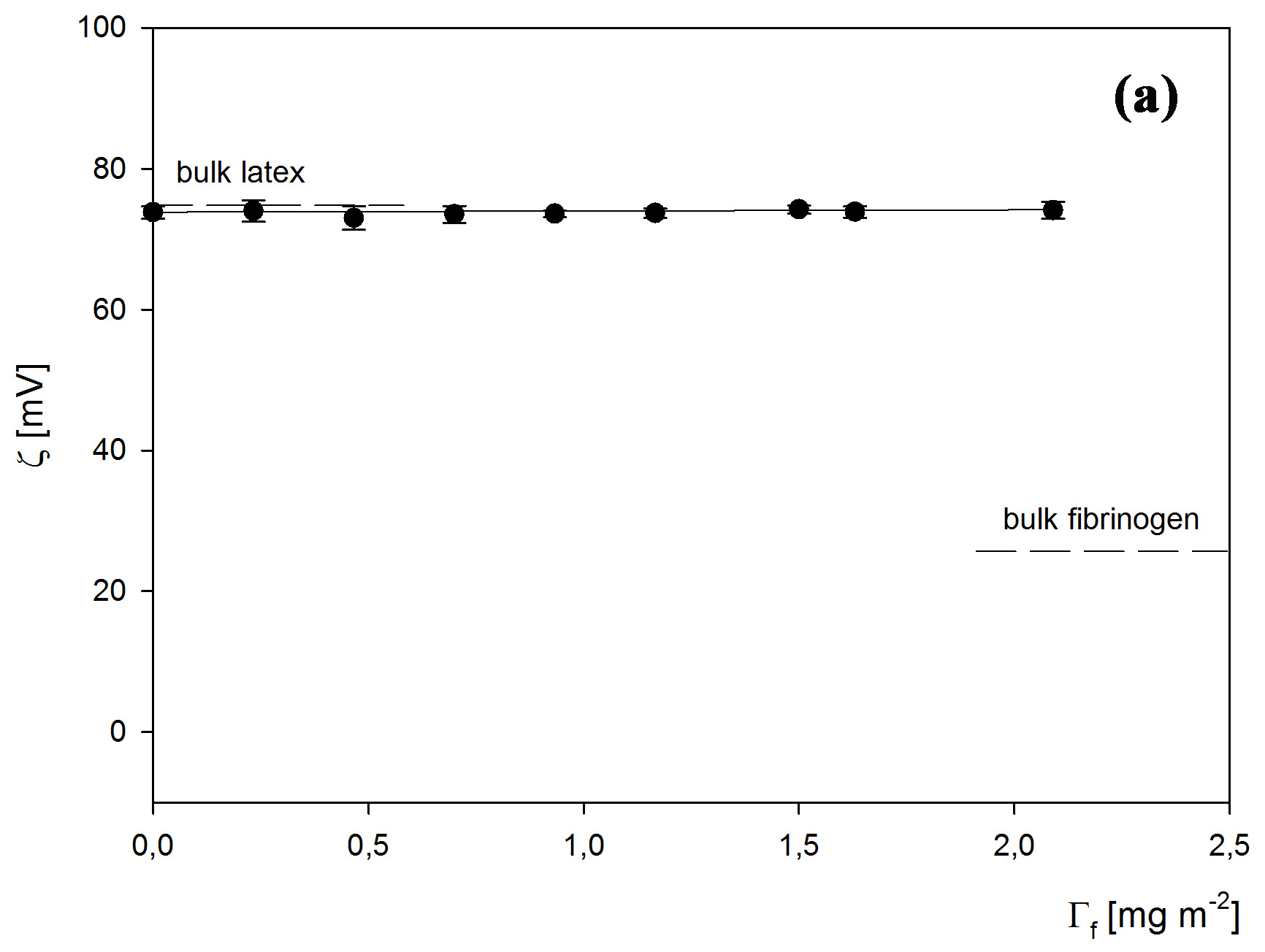} \qquad
\includegraphics[width=0.44\textwidth]{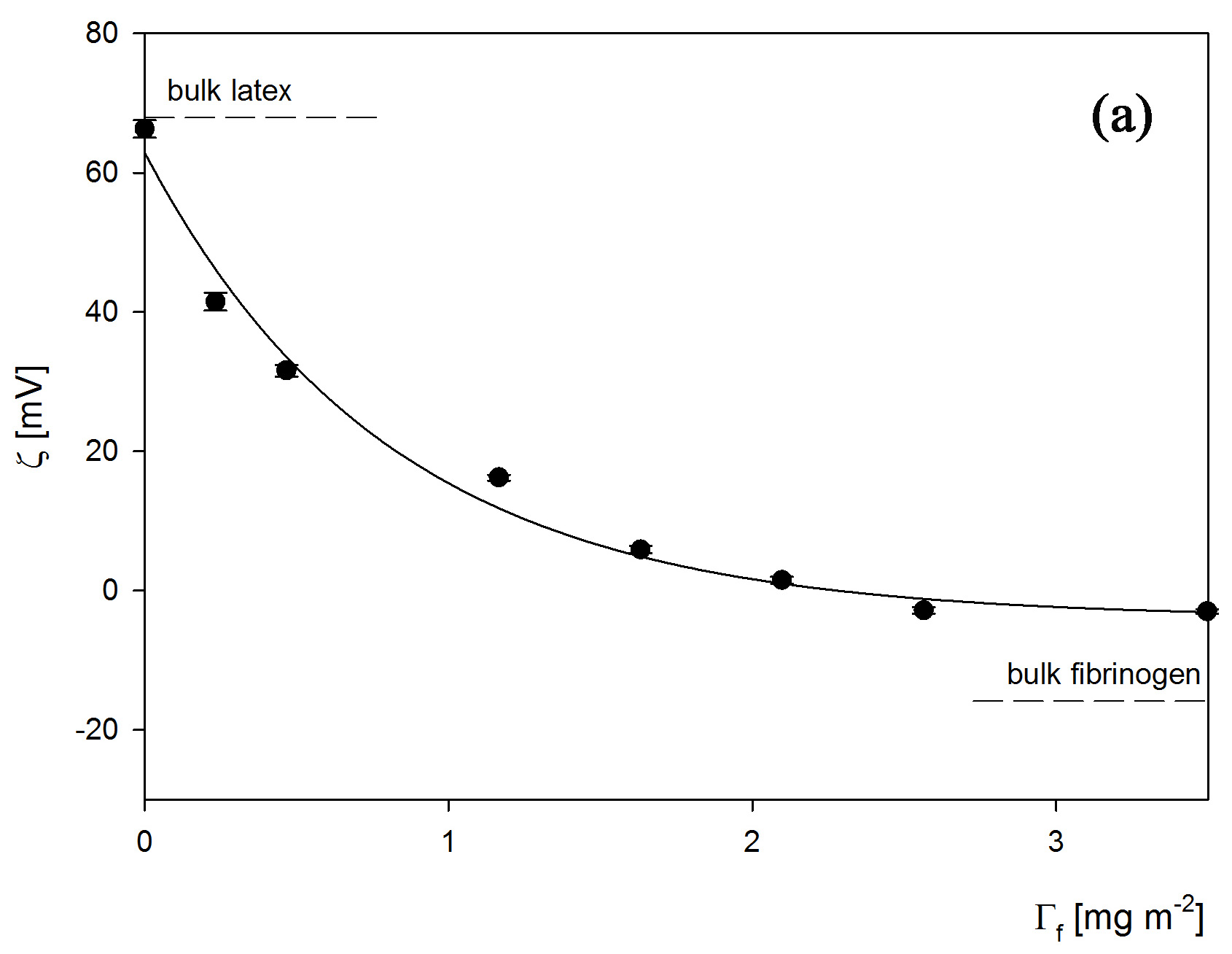}\\
\includegraphics[width=0.45\textwidth]{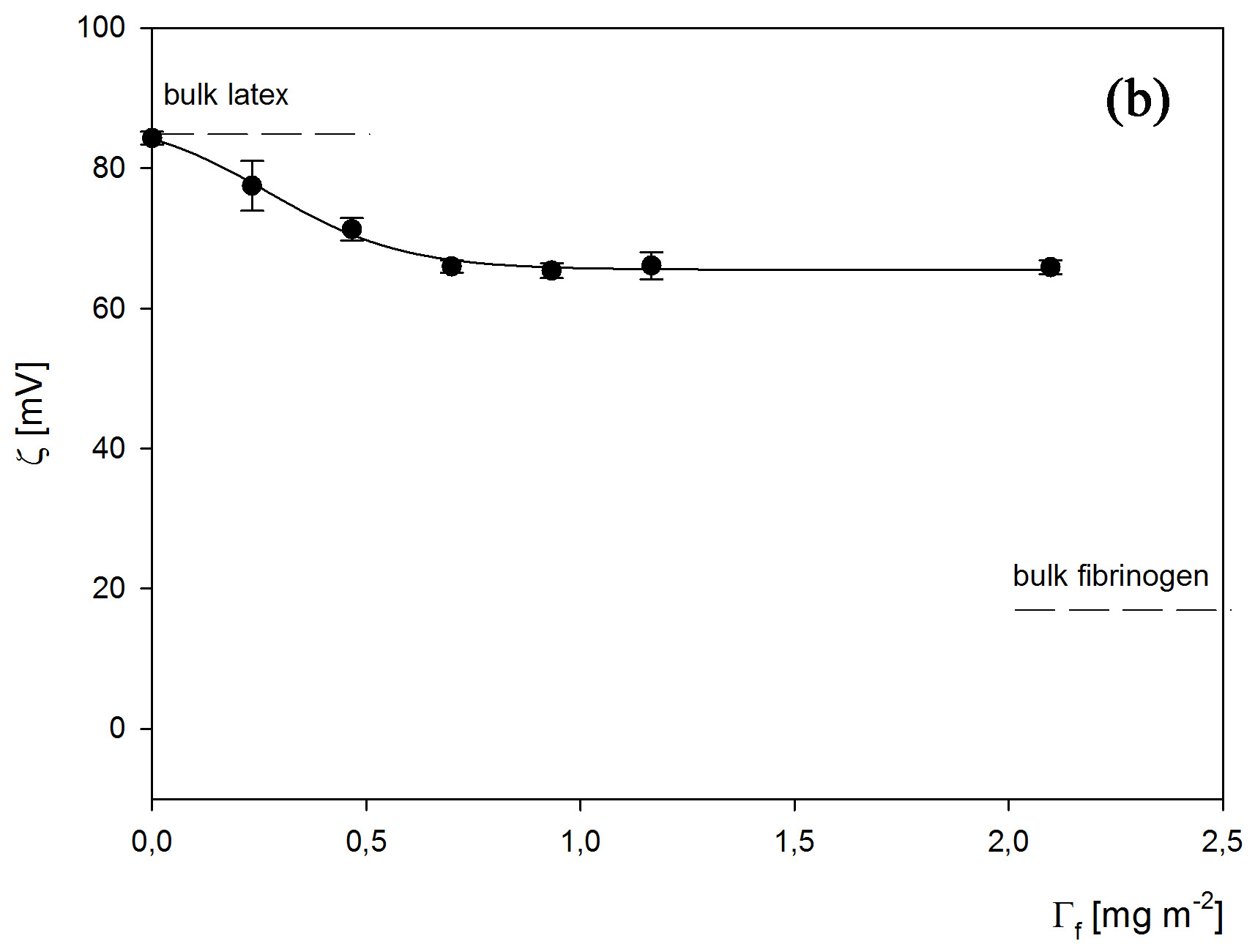} \qquad
\includegraphics[width=0.45\textwidth]{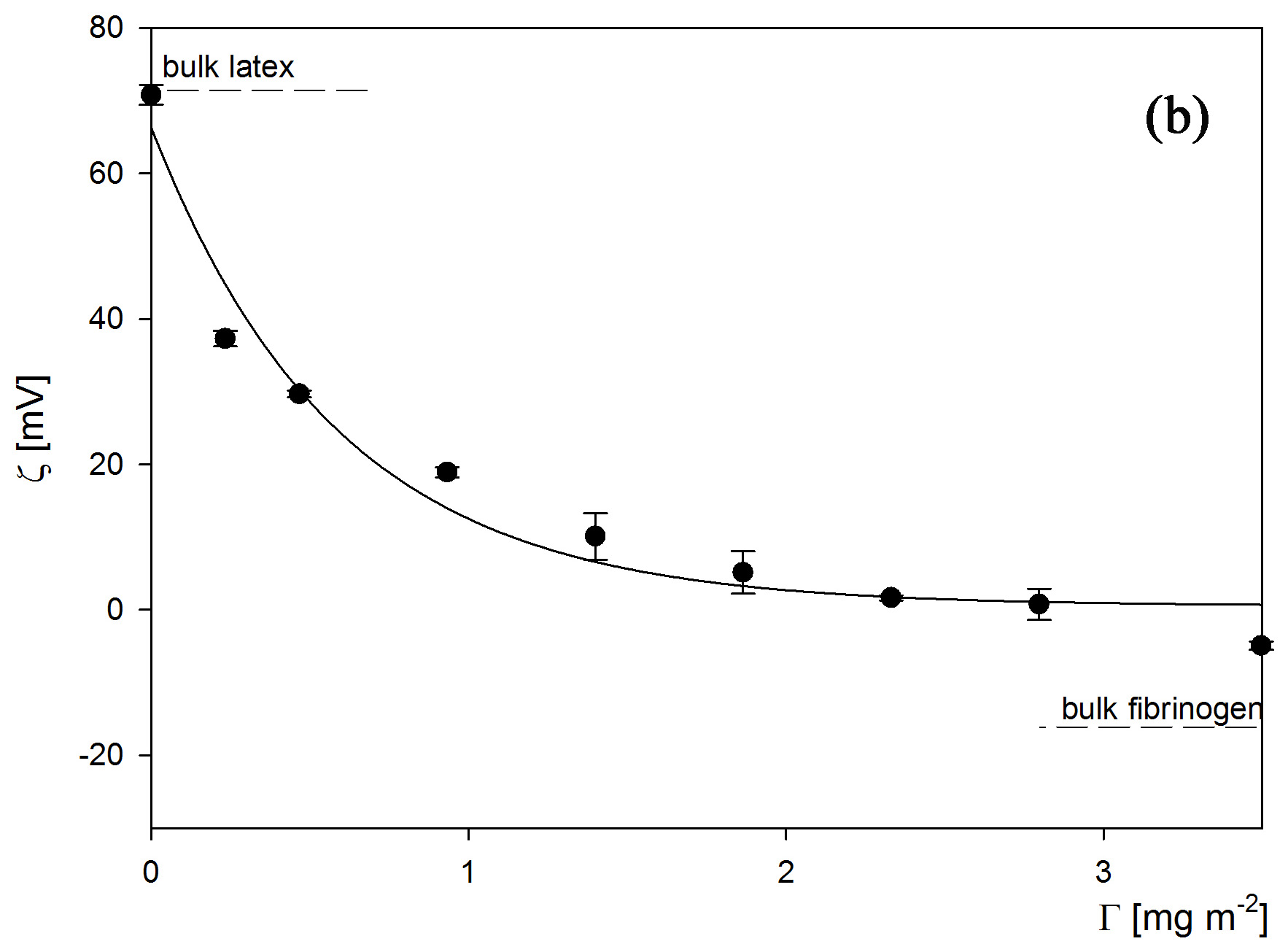} \\
\includegraphics[width=0.45\textwidth]{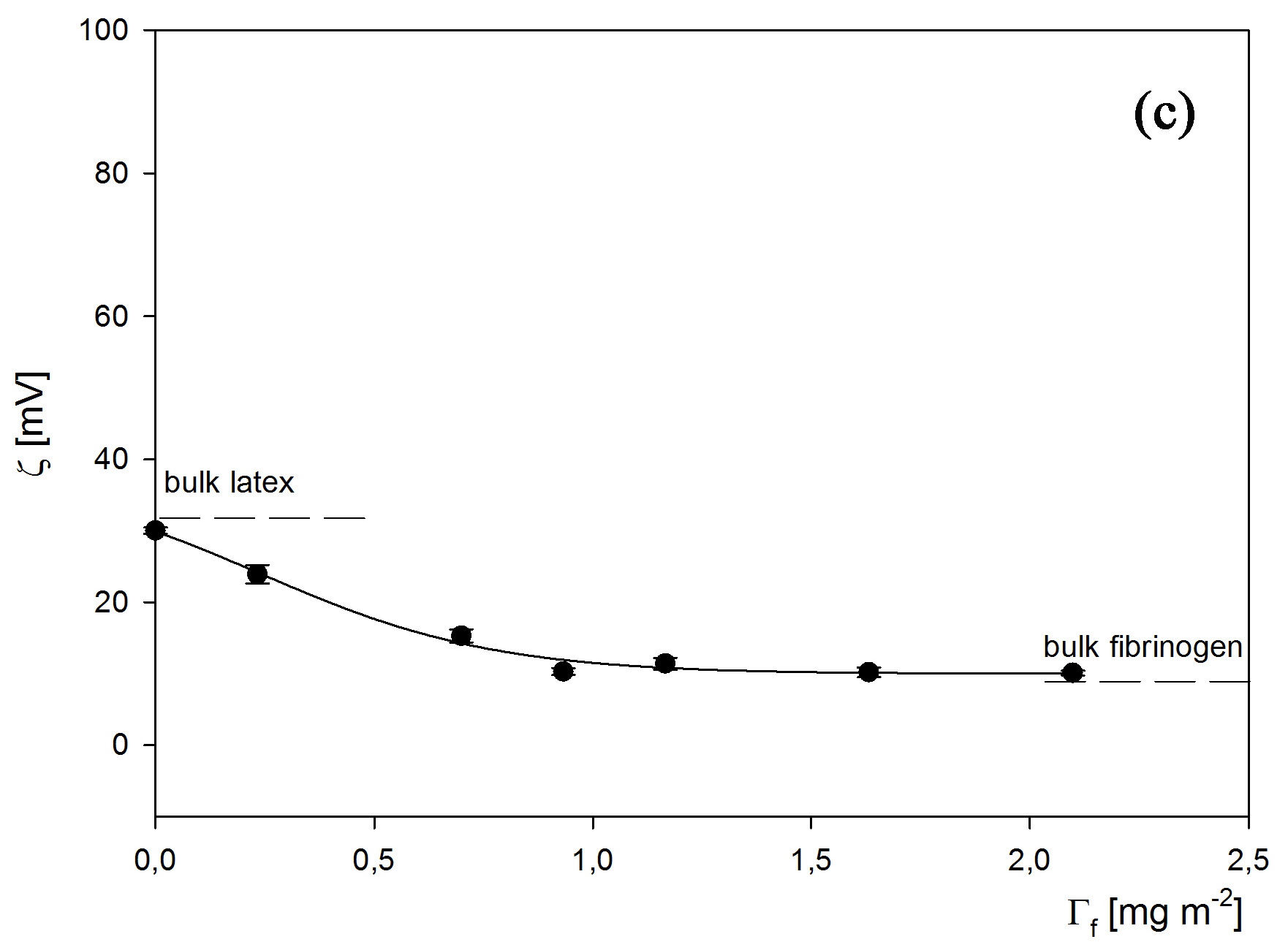} \qquad
\includegraphics[width=0.45\textwidth]{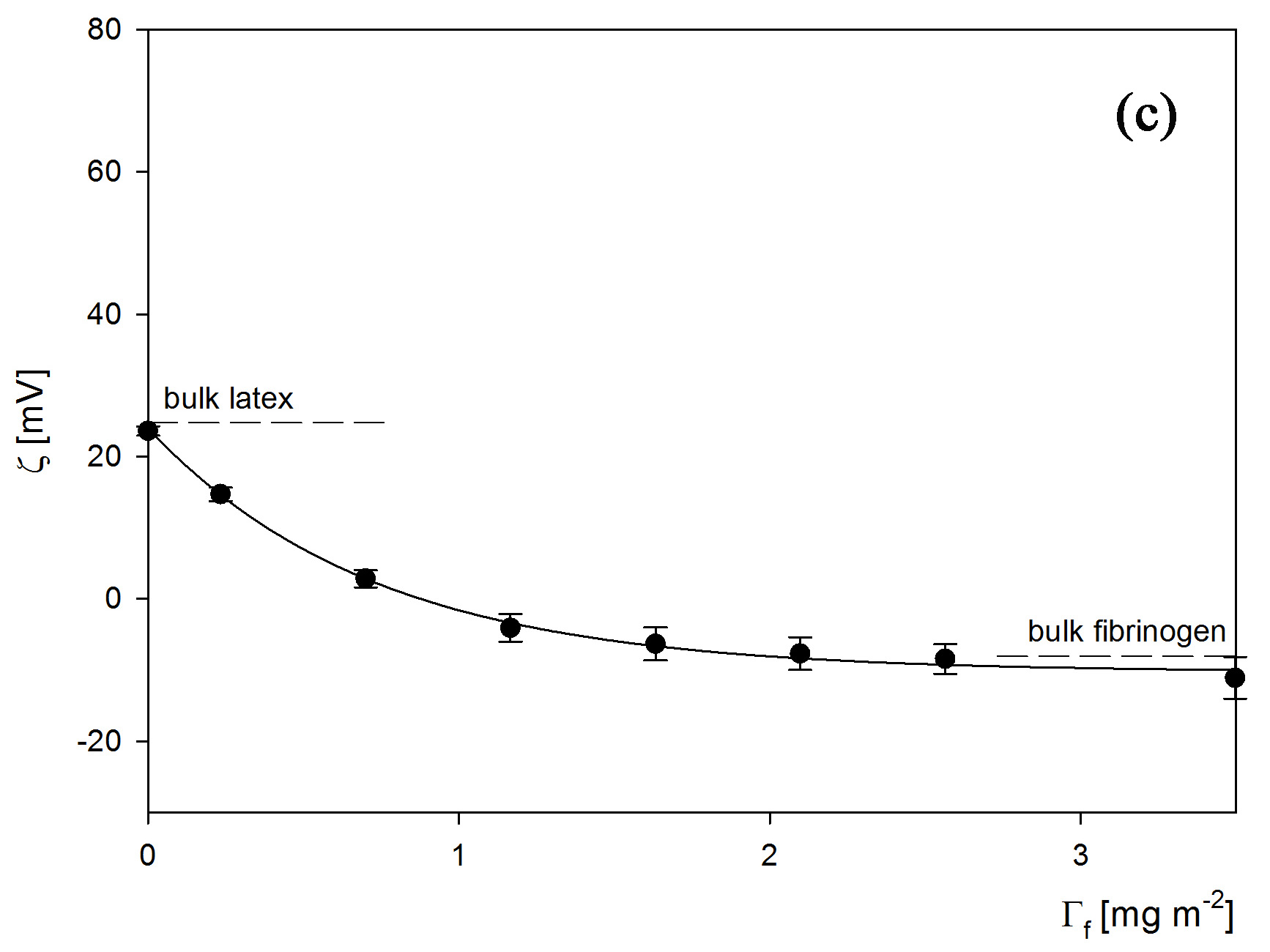} \\
\end{center}
\vspace{-7mm}
\parbox[t]{0.55\textwidth}{
\caption{The dependence of the zeta potential of latex on the nominal fibrinogen coverage. The points denote experimental results obtained for $\text{pH} = 3.5$ and various NaCl concentration: $10^{-3}$~M (a), $10^{-2}$~M (b) and 0.15~M (c). The lines are guides for the eye for the experimental data.\label{fig2}}
}
\hfill
\parbox[t]{0.45\textwidth}{
\caption{Same as in figure~\ref{fig2}, but at $\text{pH} = 9.7$.\label{fig3}}
}
\end{figure}

Experimental results expressed as the dependence of the microsphere zeta potential on the fibrinogen coverage calculated from equation (\ref{eq7}) are shown in figure~\ref{fig2} (parts a, b, c) for $\text{pH} = 3.5$, and figure~\ref{fig3} (parts a, b, c) for $\text{pH} = 9.7$ and various ionic strengths $10^{-3} - 0.15$~M NaCl. As can be seen, in figure~\ref{fig2}~(a), at $\text{pH} = 3.5$ and  $I =10^{-3}$~M, the zeta potential of latex remains constant for the entire range of the fibrinogen concentration during the adsorption experiments. This indicates that there is no fibrinogen adsorption under these conditions. However, for higher NaCl concentrations, $\zeta_\text{f}$ significantly decreases with $\Gamma_\text{f}$ attaining limiting values that are higher than the zeta potential of fibrinogen in the bulk. Note, that $\zeta_\text{f}$ becomes constant where the maximal coverage is approached. Therefore, one can roughly estimate it from the dependence of $\zeta_\text{f}$ on fibrinogen coverage on latex particles $\Gamma_\text{f}$. Analysing figure~\ref{fig2}, the limiting fibrinogen coverages (where the zeta potential does not change) are approximately 0.9 and 1.1~mg~m$^{-2}$ for NaCl concentration of $10^{-2}$, 0.15~M, respectively. In an analogous way, the limiting coverages for $\text{pH} = 9.7$ and the NaCl concentration $10^{-3}$, $10^{-2}$, 0.15~M (figure~\ref{fig3}) were estimated to be 2.0, 2.3, 3.4~mg~m$^{-2}$, respectively. These results indicate that significant adsorption of fibrinogen on latex particles occurred for the applied ionic strength.

It should be mentioned, however, that the maximum coverages of fibrinogen derived from the zeta potential measurements are of a limited precision. Therefore, in order to more exactly determine the maximum coverage, a quantitative depletion method described in reference \cite{b30} was applied. According to this procedure, the residual fibrinogen remaining in the suspension after the adsorption step is adsorbed on mica and then imaged by AFM. Fibrinogen adsorption is carried out under diffusion transport conditions using the latex suspension without applying any filtration or centrifugation. In these measurements, the average number of fibrinogen molecules per unit area of mica $N_\text{fm}$ is determined as a function of time. The principle of this method is that under diffusion-controlled transport, $N_\text{fm}$ increases proportionally to the concentration of fibrinogen, i.e., \cite{b34}
\begin{equation}
\label{eq8}
N_\text{fm} = \left( \frac{D t}{\pi} \right)^{\frac{1}{2}} c_\text{fr} = C_\text{f} c_\text{fr},
\end{equation}
where $C_\text{f}  = \left( {D t}/{\pi} \right)^{\frac{1}{2}}$ is a constant, which can be calculated if the adsorption time $t$ and the diffusion coefficient of fibrinogen $D$ are known. This constant was also determined in calibrating measurements of the protein adsorption on bare mica sheets from solutions of defined concentrations. It is also important to mention that by measuring $N_\text{fm}$ and exploiting equation (\ref{eq8}), one can uniquely calculate the unknown fibrinogen concentration in the suspension.

\begin{figure}[!t]
\centerline{
\includegraphics[width=0.55\textwidth]{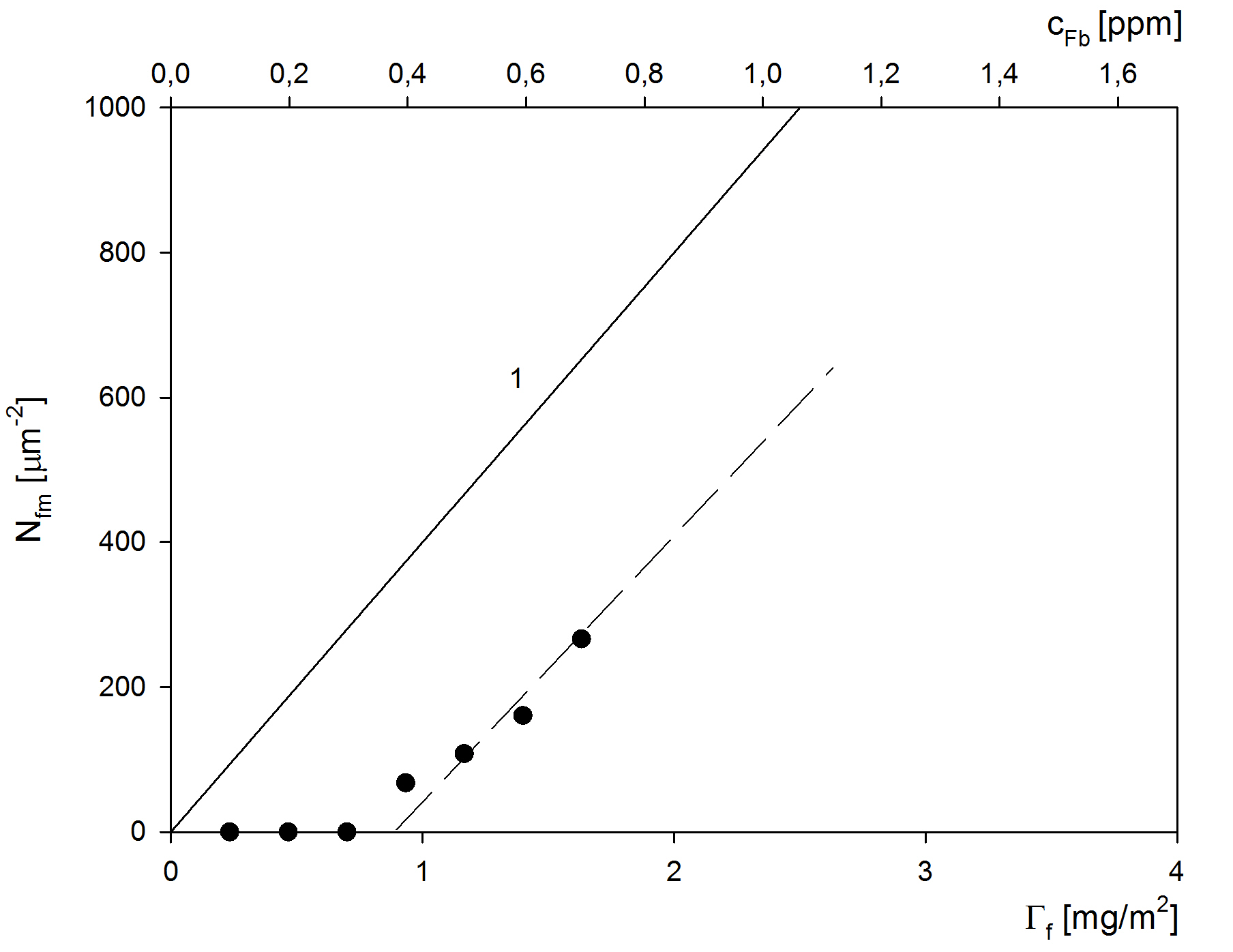}
}
\caption{The dependence of the surface concentration of fibrinogen on mica ($N_\text{fm}$) determined by AFM imaging on the fibrinogen coverage  $\Gamma_\text{f}$ (lower axis) and on the fibrinogen concentration in the suspension $c_\text{Fb}$, bulk latex concentration after mixing $c_\text{l} = 60$~mg~L$^{-1}$, $\text{pH} = 3.5$, and NaCl concentrations equal to $10^{-2}$~M. The solid line 1 represents the reference results predicted for diffusion-controlled transport of fibrinogen and the dashed line is a guide for eye for the experimental data.}
\label{fig4}
\end{figure}

\begin{figure}[!b]
\centerline{
\includegraphics[width=0.55\textwidth]{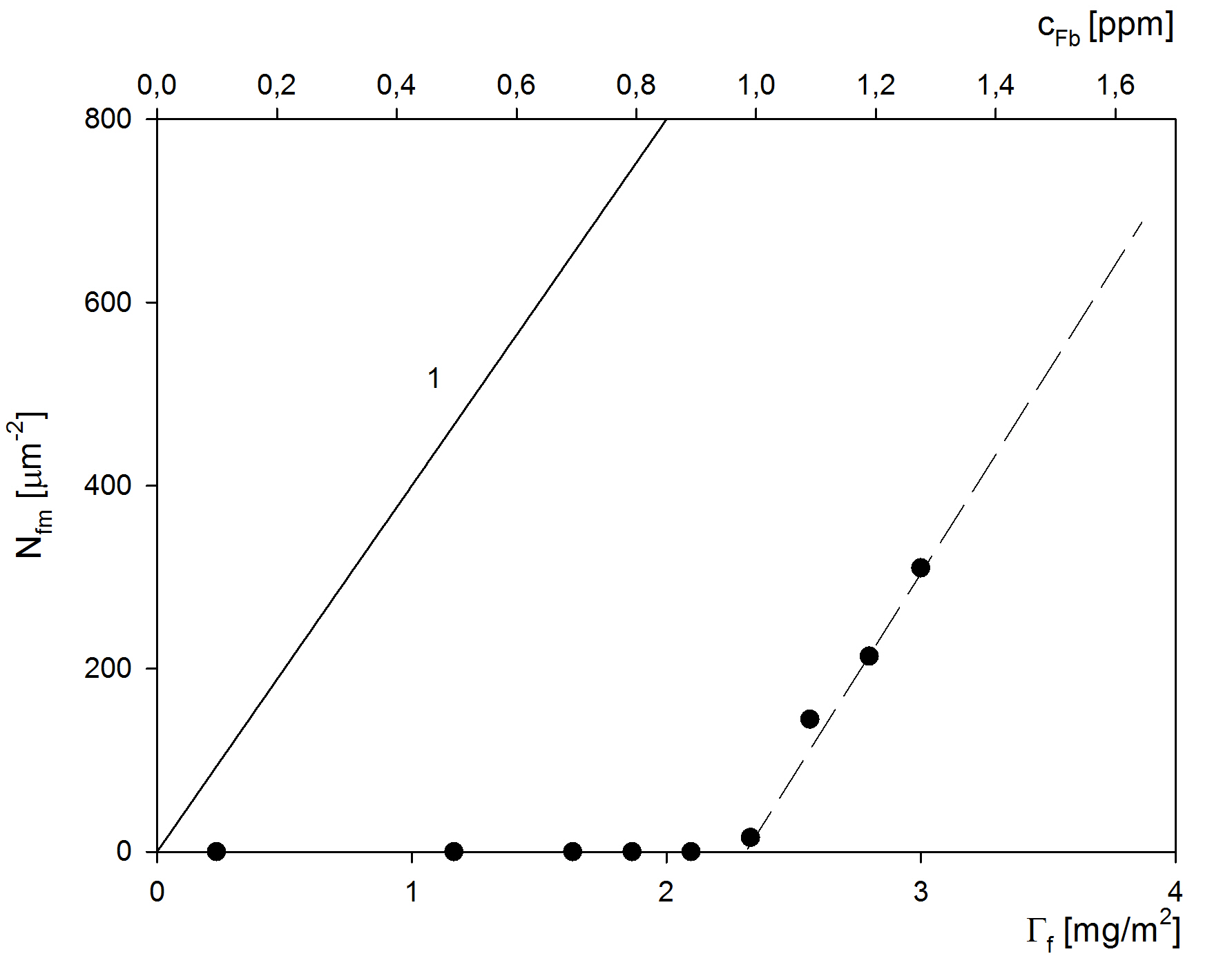}
}
\caption{Same as in figure~\ref{fig4}, but at $\text{pH} = 9.7$.}
\label{fig5}
\end{figure}

It should be mentioned that the latex particle deposition on mica during the fibrinogen adsorption runs is negligible due to their low concentrations and diffusion coefficients. The results obtained in these experiments are shown in figure~\ref{fig4} for $\text{pH} = 3.5$ and figure~\ref{fig5} for $\text{pH} = 9.7$ (NaCl concentrations equal to $10^{-2}$~M)  as the dependence of $N_\text{fm}$ on the initial concentration of fibrinogen in the suspension (the final latex particle concentration was 60~mg~L$^{-1}$). Additionally, in figures~\ref{fig4} and \ref{fig5}, the reference data for fibrinogen solutions of the same concentration without contacting the latex are plotted as straight lines number 1. As can be seen, for low fibrinogen concentration in the initial solution, there was no adsorption on mica and $N_\text{fm}$ fluctuated within error bounds near zero. Afterward, if $c_\text{Fb}$, exceeds threshold values, a liner increase in $N_\text{fm}$  is observed with slopes similar to the reference data. This confirms that fibrinogen adsorption on latex was irreversible for the pH and  ionic strength conditions presented.

By extrapolating the linear experimental dependencies shown in figures~\ref{fig4} and \ref{fig5} to zero, one can obtain the threshold fibrinogen concentrations denoted by $c_\text{fmax}$. Knowing the concentrations, the maximum coverage of fibrinogen on latex can be calculated directly from equation (\ref{eq7}) by substituting $c_\text{f}$ by $c_\text{fmax}$ In this way, one obtains precise values of $\Gamma_\text{max}$ that are listed in table~\ref{tab2}. As can be noticed, the maximum coverages agree with the previous results derived from the zeta potential measurements (see figures~\ref{fig2}--\ref{fig3}). Thus, for $\text{pH} = 3.5$, the maximum coverage is negligible for NaCl concentrations equal to $10^{-3}$~M. However, for higher NaCl concentrations of $10^{-2}$ and 0.15~M,  the maximum fibrinogen coverage assumes quite appreciable values of 0.9 and 1.1~mg~m$^{-2}$, respectively. The latter results prove that  the classical mean-field approach, where it is assumed that the effective charge on protein molecules is uniformly distributed and  the molecule is characterized by an average value of zeta potential, is inadequate. This is so, because  both the latex particles and fibrinogen molecule exhibit a large (positive) zeta potential at $\text{pH} = 3.5$ for NaCl concentration of $10^{-2}$  and 0.15~M (see table~\ref{tab1}). This discrepancy can be explained in terms of a heterogeneous charge distribution over the fibrinogen molecule with a negative charge located in  its core part whereas the rest of the molecule remains positively charged. This assumption is supported by the fact that the core part of the fibrinogen molecule comprises aspartic and glutamic acids residues characterized by pH value below 3 \cite{b36}.

\begin{table}[!h]
\caption{Maximum coverage of fibrinogen on positively and negatively charged latex particles expressed in mg~m$^{-2}$ for various pHs and NaCl concentrations.
\label{tab2}}
\vspace{2ex}
\begin{center}
\begin{tabular}{|c|c|c|c|c|c|}
\hline\hline
\multirow{3}{*}{pH} & NaCl & \multicolumn{4}{|c|}{Maximum coverage of fibrinogen on latex [mg m$^{-2}$]}  \\
\cline{3-6}
 & concentration & \multicolumn{2}{|c|}{$\Gamma_\text{max}$ } & \multicolumn{2}{|c|}{ theoretical (RSA modelling) } \\
\cline{3-6}
& [M] & negative latex$^*$ & positive latex &  & remarks \\
\hline
\hline
\multirow{3}{*}{3.5}
 & 10$^{-3}$ & 1.8\,$\pm$\,0.2 & < 0.1             & negligible & side-on adsorption \\
\cline{2-6}
 & 10$^{-2}$ & 2.5\,$\pm$\,0.2 & 0.9\,$\pm$\,0.1 & 1.13        &	side-on adsorption \\
\cline{2-6}
 & 0.15          & 3.6\,$\pm$\,0.2 & 1.1\,$\pm$\,0.1 &  1.22        & side-on adsorption \\
\hline
\multirow{3}{*}{7.4}
 & $10^{-3}$ & 1.9\,$\pm$\,0.2 & 0.6\,$\pm$\,0.1$^*$ & 0.65 & side-on adsorption \\
\cline{2-6}
 & 10$^{-2}$ & 2.7\,$\pm$\,0.2 & 1.2\,$\pm$\,0.1$^*$ & 1.12 & side-on adsorption \\
\cline{2-6}
 & 0.15          & 3.2\,$\pm$\,0.2 & 1.3\,$\pm$\,0.1$^*$ & 1.29 & side-on adsorption \\
\hline
\multirow{3}{*}{9.7}
 & 10$^{-3}$ & -- & 2.0\,$\pm$\,0.2 & 1.8 & unoriented adsorption \\
\cline{2-6}
 & 10$^{-2}$ & -- & 2.3\,$\pm$\,0.2 & 2.8 & unoriented adsorption \\
\cline{2-6}
 & 0.15         & -- & 3.4\,$\pm$\,0.2  & 3.7 & unoriented adsorption \\
\hline\hline
\end{tabular}
\end{center}
\qquad \small{$^*$Previous results obtained in references \cite{b30,b31,b32}.}
\end{table}

Therefore, our experimental studies show that even at $\text{pH} = 3.5$, the fibrinogen molecule contains negatively charged fragments which is in conflict with the previous reports where it was assumed that the entire molecule is positively charged at this pH \cite{b24,b34,b43}.

\begin{figure}[!t]
\begin{center}
\includegraphics[width=0.64\textwidth]{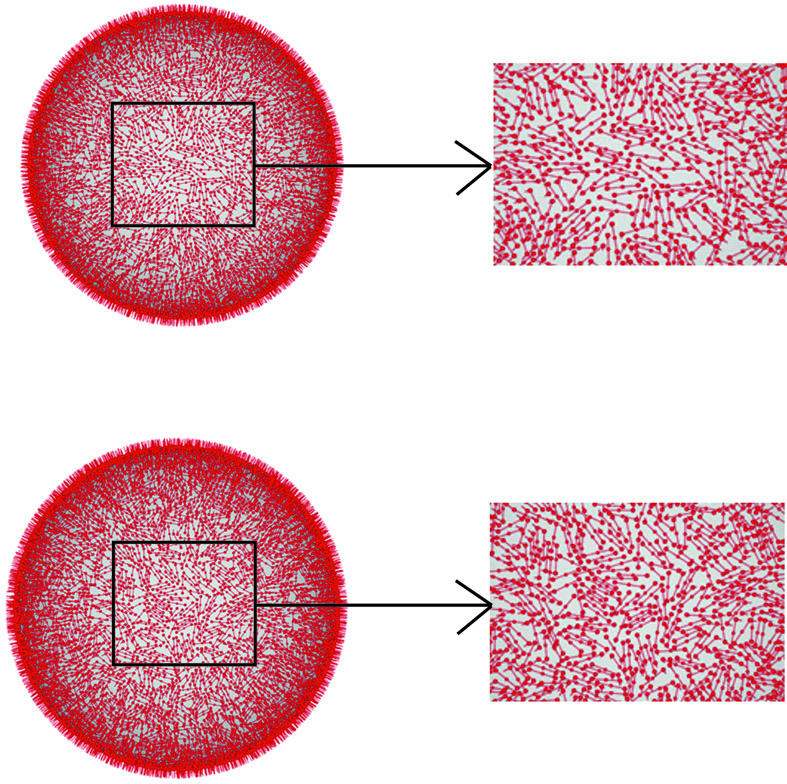}
\end{center}
\caption{(Color online) The topology of fibrinogen molecules (approximated  by the Bead Model B) on the positive latex particles derived from the RSA simulations  at $\text{pH} = 3.5$ and for different electrolyte concentrations: top: $10^{-2}$~M,  1.13~mg~m$^{-2}$ and  bottom: 0.15~M, 1.22~mg~m$^{-2}$.}
\label{fig6}
\end{figure}

The presence of a negative charge at the core part of the molecule suggests that fibrinogen adsorption at $\text{pH} = 3.5$ occurs exclusively in the side-on orientation with the $\alpha$C domains of the A$\alpha$ chains pointing toward the bulk solution. This assumption is supported by the fact that the theoretically predicted coverage derived by applying the random sequential adsorption modelling \cite{b30} is equal to 1.13 and 1.22~mg~m$^{-2}$  for NaCl concentration of $10^{-2}$ and 0.15~M, which is slightly lower than with  experimental values.  Snapshots of fibrinogen monolayers on latex particles derived from these simulations are shown in figure~\ref{fig6} for the ionic strength $10^{-2}$ and 0.15~M NaCl.

As can be seen, it is theoretically predicted that fibrinogen molecules adsorb in the side-on orientation exposing the positively charged side arms into the electrolyte solution. Additionally, due to the lateral electrostatic repulsion among the beads forming the core part of molecules, the monolayer obtained for 10$^{-2}$~M, NaCl is characterized by a significantly smaller density compared to the jamming coverage for hard (non-interacting) spheres predicted to be 1.29~mg~$^{-2}$ \cite{b24}.

For $\text{pH} = 7.4$, the experimental maximum coverages are 1.2 and 1.3~mg~m$^{-2}$  that are only slightly larger than at $\text{pH} = 3.5$ but considerably lower than for fibrinogen adsorption at negatively charged latex determined in reference \cite{b32} and equal to 2.7 and 3.2~mg~m$^{-2}$. This seems to be paradoxical since at $\text{pH} = 7.4$, the net charge of the fibrinogen molecule is strongly negative (see table~\ref{tab1}). In the case of positively charged microspheres, this discrepancy can be explained by the appearance of a strong electrostatic attraction between the particle surface and fibrinogen molecules promoting their side-on orientation.

	However, at $\text{pH} = 9.7$, where the fibrinogen molecule charge is considerably more negative, the maximum coverages significantly increase attaining 2.0, 2.3 and 3.4~mg~m$^{-2}$ for NaCl concentration of $10^{-3}$, $10^{-2}$ and 0.15~M, respectively. As can be noticed, (see table~\ref{tab2}), these results are similar to  the previously reported for the fibrinogen adsorption at negatively charged latex at $\text{pH} = 3.5$ \cite{b30}. They were interpreted in terms of the RSA model assuming unoriented adsorption of fibrinogen, i.e., simultaneously occurring in the side-on and end-on orientations that prevails for long adsorption times. Additionally, in these calculations the lateral electrostatic repulsion among the adsorbed fibrinogen molecules was considered to be responsible for the decrease of the maximum coverage for a lower NaCl concentration (see the last column in table~\ref{tab2}). Therefore, the agreement of our results obtained for positive latex with these theoretical predictions confirms that at $\text{pH} = 9.7$, fibrinogen also adsorbs at positively charged latex in the end-on orientation.

\subsection{Stability of fibrinogen monolayers on latex particles}

\begin{figure}[!t]
\centerline{
\includegraphics[width=0.55\textwidth]{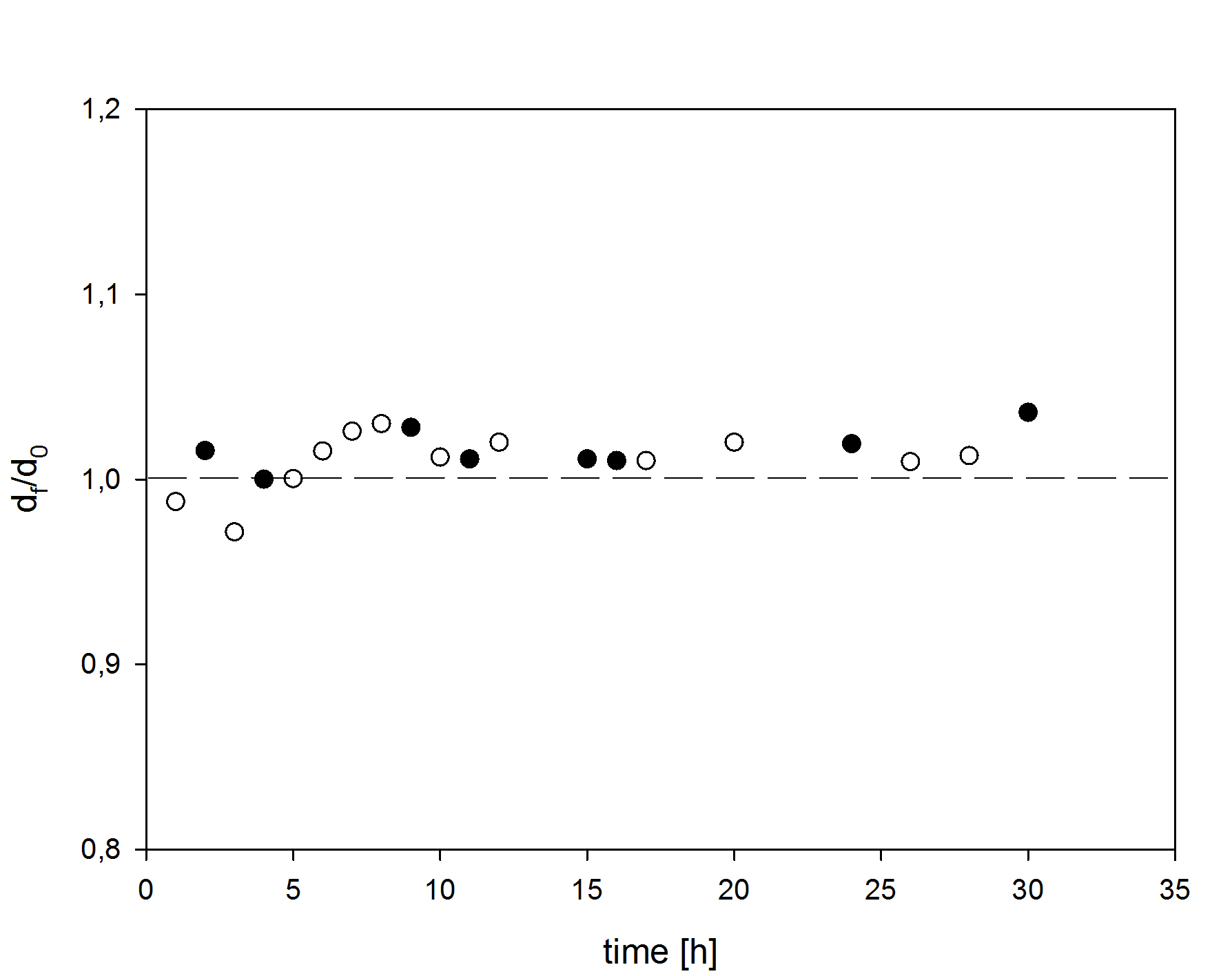}
}
\caption{The dependence of normalized hydrodynamic diameter of fibrinogen $d_\text{f}/d_0$ on time for  $I=10^{-2}$~M NaCl, $\text{pH} = 3.5$, $\Gamma_\text{f} = 0.9$~mg~m$^{-2}$ and $\Gamma_\text{f} =  2.0$~mg~m$^{-2}$ for white and black dots, respectively.}
\label{fig7}
\end{figure}

In these series of experiments, the stability of fibrinogen monolayers on  microspheres  was checked by measuring their  hydrodynamic diameter as function of time and in pH cycling experiments. The experimental results are shown in figure~\ref{fig7} as the dependence of $d_\text{f}/d_0$ ($d_0$ is the initial hydrodynamic diameter of fibrinogen-covered latex) on the storage time for NaCl concentration equal to $10^{-2}$~M and $\text{pH} = 3.5$.

As can be seen in figure~\ref{fig7},  there were  no significant changes in the normalized hydrodynamic diameter for the time period up to 35~h.

\clearpage

Additionally, the stability of fibrinogen monolayers was checked in pH cycles. This experiment consisted of the following steps:
\begin{itemize}

\item[(i)] a fibrinogen monolayer of a well-defined coverage was adsorbed on latex particles at $\text{pH} = 3.5$ and NaCl concentration of 0.15~M,

\item[(ii)] pH was stepwise changed by the addition of HCl (pH decrease) or NaOH (pH increase) by keeping the changes in the initial ionic strength negligible

\item[(iii)] after stabilization of pH,  the electrophoretic mobility of latex was measured

\item[(iv)] the entire process was repeated three times in the range of pH between 3.5 and 9.7.

\end{itemize}

\begin{figure}[!b]
\centerline{
\includegraphics[width=0.6\textwidth]{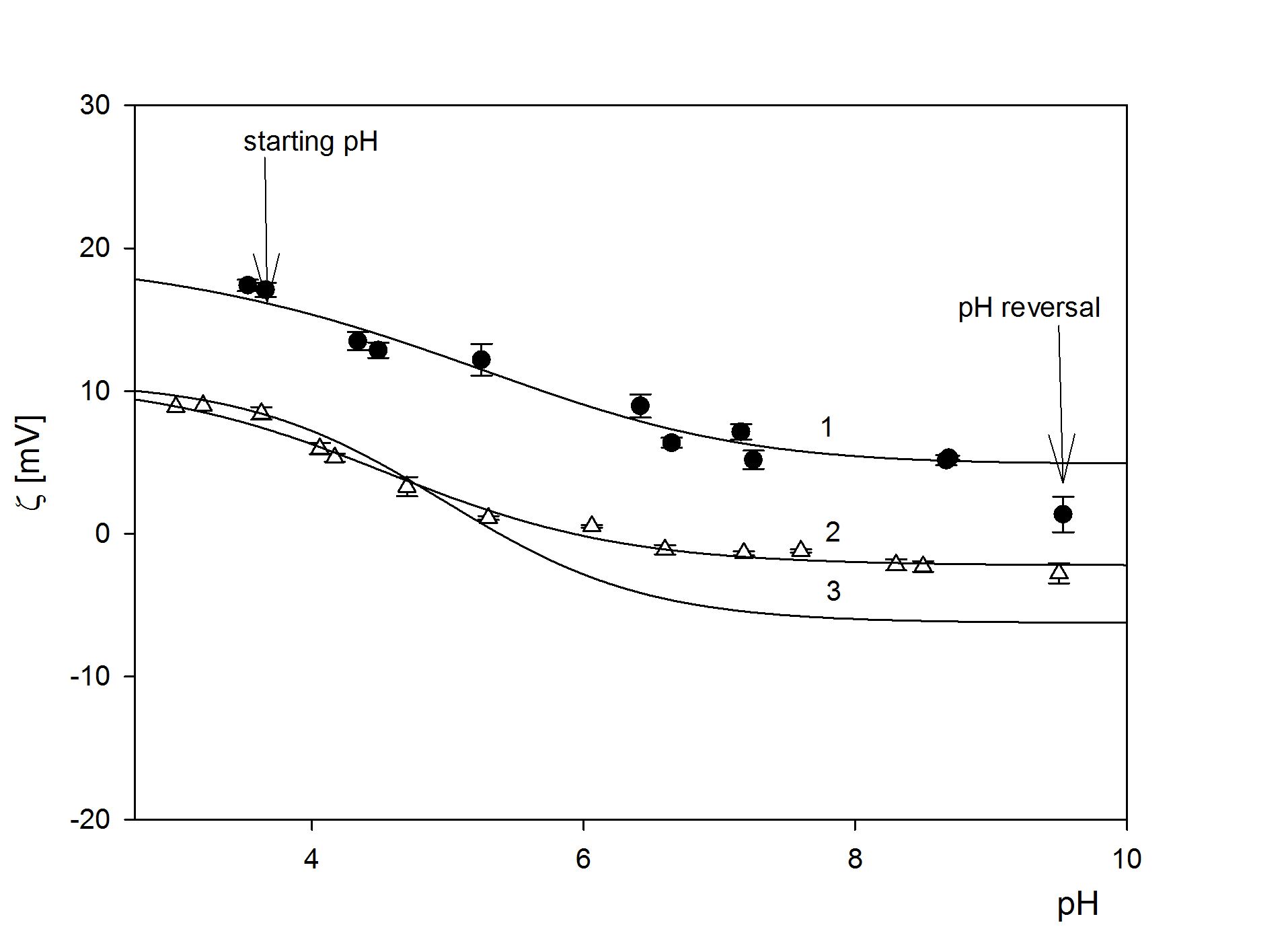}
}
\caption{The dependence of fibrinogen monolayers adsorbed at $\text{pH} = 3.5$, $I=0.15$~M NaCl, on pH cycling starting from 3.5 to 9.7 and back to 3.5. The points denote experimental results for: $\Gamma_\text{f} = 2$~mg~m$^{-2}$ (triangles), $\Gamma_\text{f} = 0.9$~mg~m$^{-2}$ (dots). Lines 1 and 2 are fits of experimental data, and line 3 denotes the reference results for fibrinogen zeta potential in the bulk.}
\label{fig8}
\end{figure}

Experimental results obtained in these measurements are shown in figure~\ref{fig8}. It can be observed that under these conditions, no irreversible conformational changes in the protein monolayers were observed.


\section{Conclusions}

The AFM and micro-electrophoretic measurements complemented by extensive RSA modelling delivered  reliable information about  the mechanisms of fibrinogen adsorption on positively charged microspheres for a broad range of pHs and ionic strength. It was confirmed that fibrinogen adsorption was  irreversible and the monolayer properties were  stable in time.

Interesting results were obtained at $\text{pH} = 3.5$, where for NaCl concentrations of $10^{-2}$ and 0.15~M,  the maximum fibrinogen coverage of fibrinogen on microspheres was  quite significant amounting to 0.9 and 1.1~mg~m$^{-2}$, respectively. This contradicts the predictions derived from the classical mean-field approach because both the fibrinogen molecules and the latex particles exhibit a positive zeta potential at this pH. This discrepancy was explained in terms of a heterogeneous charge distribution over the fibrinogen molecule with a negative charge located at its core part whereas the rest of the molecule remains positively charged. This anomalous adsorption of fibrinogen confirms that, in contrast to common views, a negative charge is present at the body of the fibrinogen molecule even at such low pH. This enables a side-on adsorption of the molecules interpreted by the random sequential adsorption modelling.

The increase in the fibrinogen coverage with NaCl concentration was interpreted as due to the electrostatic repulsion among the adsorbed protein molecules.
On the other hand, for $\text{pH} = 9.7$, the maximum coverage of fibrinogen on latex  was much larger, varying between 2.0~mg~m$^{-2}$ and 3.4~mg~m$^{-2}$ for $10^{-3}$ and 0.15~M NaCl, respectively, which almost fully matches the previous results obtained for negative latex at $\text{pH} = 3.5$.  These results were in agreement with the theoretical RSA modelling carried out by assuming the side-on and end-on orientations prevailing at longer times. This confirmed  a different adsorption mechanism of fibrinogen on latex at $\text{pH} = 9.7$, characterized by the presence of a considerable amount of end-on adsorbed molecules. The increase in the coverage with NaCl concentration was interpreted in terms of the diminishing range of the repulsive electrostatic interactions among the adsorbed molecules.

Besides a significance to basic science, our results have practical implications for developing a robust procedure of preparing stable fibrinogen monolayers at latex particles of a well-controlled coverage and orientation of molecules. Such latex/fibrinogen complexes can be exploited to study interactions with antibodies and other ligands.

\section*{Acknowledgements}
This work was financially supported by the NCN Grants: PRELUDIUM-2013/09/N/ST4/00320 and    UMO-2012/07/B/ST4/00559.

\clearpage

\ukrainianpart

\title
{Моделювання та вимірювання фібриногенної адсорбції \\ на додатньо заряджених мікросферах}
\author{П. Зелішевска\refaddr{label1},
A. Братек-Скіцький\refaddr{label1}, З. Адамчик\refaddr{label1}, M. Цєсла\refaddr{label2}}
\addresses{
\addr{label1}  Інститут каталізу і хімії поверхні ім. Я. Габера Польської академії наук, Краків, Польща
\addr{label2}  Інститут фізики ім. М. Смолуховського, Ягелонський університет,  Краків, Польща
}

\makeukrtitle

\begin{abstract}
Теоретично та експериментально досліджено адсорбцію фібриногену на позитивно заряджених мікросферах.
Структура моношарів і максимальне покриття були визначені експериментально при  $\text{pH} = 3.5$ і 9.7 для концентрації NaCl в діапазоні $10^{-3} - 0.15$~M. Максимальне покриття фібриногену на частинках латексу було визначене прецизійно методом мікроскопу атомної сили (AFM). Несподівано, що при $\text{pH} = 3.5$, коли і фібриноген і частинки латексу заряджені позитивно, максимальне покриття змінюється між 0.9~мг~м$^{-2}$ і 1.1~мг~м$^{-2}$ для $10^{-2}$ і 0.15~M NaCl, відповідно. При $\text{pH}=9.7$ максимальне покриття фібриногену є більшим і змінюється від 1.8~мг~м$^{-2}$ до 3.4~мг~м$^{-2}$ для $10^{-2}$ і 0.15~M NaCl, відповідно. Експериментальні результати підтверджені чисельним моделюванням.
\keywords фібриногення адсорбція, позитивно заряджені мікросфери

\end{abstract}

\end{document}